# The Peculiar Velocity Function of Galaxy Clusters

Neta A. Bahcall

and

Siang Peng Oh

Princeton University Observatory, Princeton NJ 08544

# ABSTRACT

The peculiar velocity function of clusters of galaxies is determined using an accurate sample of cluster velocities based on Tully-Fisher distances of Sc galaxies (Giovanelli *et al* 1995b). In contrast with previous results based on samples with considerably larger velocity uncertainties, the observed velocity function does not exhibit a tail of high velocity clusters . The results indicate a low probability of $\lesssim 5\%$ of finding clusters with one-dimensional velocities greater than $\sim 600$ km s$^{-1}$. The root-mean-square one-dimensional cluster velocity is 293±28 km s$^{-1}$. The observed cluster velocity function is compared with expectations from different cosmological models. The absence of a high velocity tail in the observed function is most consistent with a low mass-density ($\Omega \sim 0.3$) CDM model, and is inconsistent at $\gtrsim 3\sigma$ level with $\Omega = 1.0$ CDM and HDM models. The root-mean-square one-dimensional cluster velocities in these models correspond, respectively, to 314, 516, and 632 km s$^{-1}$ (when convolved with the observational uncertainties). Comparison with the observed RMS cluster velocity of 293±28 km s$^{-1}$ further supports the low-density CDM model.

*Subject headings:* galaxies:clusters:general — cosmology:observations — cosmology:theory — dark matter — large-scale structure of universe

## 1. Introduction

The motions of clusters of galaxies can place strong constraints on cosmological models and on the mass-density of the universe. Bahcall *et al* (1994a,b), Cen *et al* (1994), Croft & Efstathiou(1994), Lauer & Postman(1994), Gramann *et al* (1995), and Moscardini *et al* (1995), showed that clusters of galaxies provide a particularly efficient and accurate way to trace the peculiar velocity field in the universe. In turn, the peculiar velocity field, caused by the gravitational growth of structure, sheds light on the cosmology responsible for the formation and evolution of the structure (Dekel 1994, Strauss & Willick 1995).

Bahcall *et al* (1994a,b) investigated the probability distribution function of cluster peculiar velocities, i.e., the cluster velocity function (CVF), and showed that it provides an important tool for distinguishing between different cosmological models. They determined the cluster velocity function for several cosmological models using large scale N-body simulations. They also determined the observed CVF using the available data and compared it with model expectations. However, the large uncertainties of the cluster velocity data broadened the CVF and produced an artificial tail of high velocity clusters. These uncertainties did not allow an accurate determination of the true underlying cluster velocity function, nor an accurate comparison with the cosmological models (since the convolution of the model CVFs with large observational uncertainties reduced the differences between the various models). Similar results were also obtained by Croft & Efstathiou(1994) and Moscardini *et al* (1995). Bahcall *et al* (1994b) concluded that a cluster sample with considerably improved velocity accuracy is needed before an accurate cluster velocity function, one that is not dominated by velocity errors, can be determined.

In this paper, we use a new sample (Giovanelli *et al* 1995a,b) of cluster velocities that has considerably higher accuracy and uniformity than previously used samples. The new sample is based on well calibrated Tully-Fisher distance indicators of Sc galaxies. We use these data to determine the cluster velocity function and compare it with expectations from cosmological models.

## 2. The Peculiar Velocity Function of Clusters

### 2.1. Model Expectations

The peculiar velocity function of clusters of galaxies represents the probability distribution of cluster velocities relative to a comoving cosmic frame. The integrated velocity function, P(>v), represents the relative number density of clusters with peculiar velocities larger than v (where v is the three-dimensional cluster motion relative to the cosmic frame). The differential velocity function, P(v), represents the relative number density of clusters with peculiar velocities in the range v±dv, per unit dv, as a function of v. The cluster velocity functions P(v) and P(>v) were determined for four cosmological models by Bahcall *et al* (1994b) using large scale N-body simulations. We use these results below.

The cosmological models investigated and their parameters are summarized in Table 1. These parameters include the matter density, $\Omega$; the cosmological constant contribution, $\Omega_\Lambda$; the Hubble constant (in units of $H_o = 100h$ km s$^{-1}$ Mpc$^{-1}$); and the normalization of the mass fluctuations on a $8h^{-1}$ Mpc scale, $\sigma_8$. The models are normalized to the large-scale microwave background anisotropy measured by COBE (Smoot *et al* 1992). (The HDM model normalization is $\sim$ 20% higher than the $\Omega$= 1 CDM on large scales). We next describe briefly the simulations that are used to represent the cosmological models.

A large-scale particle-mesh code with box size of 800 $h^{-1}$ Mpc is used to simulate the evolution of the dark matter in the models. A large simulation box is needed in order to ensure that contributions to velocities from waves larger than the box size are small, and to minimize uncertainties due to fluctuations in the small number of large waves. The simulation box contains $500^3$ cells and $250^3 = 10^{7.2}$ dark matter particles. The spatial resolution is 1.6 $h^{-1}$ Mpc. (A higher resolution [0.8 $h^{-1}$ Mpc] smaller box [400 $h^{-1}$ Mpc] was also studied for comparison). For more details of the simulations see Bahcall *et al* (1994b).

Clusters were selected in each simulation using an adaptive linkage algorithm. The cluster mass thresholds correspond to the observed number density of typical rich clusters as well as of groups. A total of $\sim$3000 rich clusters and $\sim$5 x $10^4$ groups were obtained in each of the simulated models. The three-dimensional and one-dimensional peculiar velocity of each cluster or group, relative to the comoving cosmic frame, was obtained from the simulation and used to determine the velocity function of groups and clusters. The simulation results are consistent with expectations from linear theory (Bahcall *et al*, 1994a,b). The clusters selected for comparison with the present data correspond to the group selection threshold, which represents the best match to the threshold of the observed groups and clusters in the current sample. The results, however, are insensitive to the exact richness threshold of the clusters (Bahcall *et al* 1994b).

The cluster velocity functions of the four models are presented in Figures 3–4 and 9–11 of Bahcall *et al* (1994b); these functions represent the "exact" CVFs (in $v_{3D}$ and $v_{1D}$), unconvolved with any observational uncertainties. The results illustrate that the differences among the four models are most apparent at the high velocity end, where the low-density models predict considerably smaller peculiar velocities than the $\Omega = 1$ models. For example, while the $\Omega = 0.3$ CDM and PBI models yield $\sim$ 5% of clusters with velocities $v_{1D} > 500$ km s$^{-1}$ and $> 800$ km s$^{-1}$ respectively, the $\Omega = 1$ CDM and HDM models exhibit $\sim$5% of clusters with high velocities of $v_{1D} > 1000$ km s$^{-1}$ and $> 1300$ km s$^{-1}$, respectively. Similarly, the root-mean-square peculiar velocity of clusters differs significantly among the models. The $\Omega=0.3$ CDM model yields the lowest RMS velocity, $< v_{1D}^2 >^{1/2} \simeq 268$ km s$^{-1}$, while the $\Omega$=1 models yield the highest velocities, $<v_{1D}^2>^{1/2} \simeq$ 500–600 km s$^{-1}$ (unconvolved with observational uncertainties). The results are summarized in Table 3 of Bahcall *et al* (1994b). The sensitivity of the CVF to the cosmology makes it a powerful tool in constraining cosmological models. We use this tool below.

## 2.2. Observations

The first determination of the cluster velocity function was made by Bahcall *et al* (1994b) who used observations of cluster velocities based on Tully-Fisher (TF) and $D_n - \sigma$ distance indicators (with data from Aaronson *et al* 1986, Faber *et al* 1989, Mould *et al* 1991, 1993, and Mathewson, Ford & Buchhorn 1992). They found a velocity function that exhibits a large tail of high velocity clusters up to $v_{1D} \sim$ 2000 km s$^{-1}$ (Figs. 10-11 of Bahcall *et al* 1994b). However, the observational uncertainties of the cluster velocities were very large, reaching $\sim$900 km s$^{-1}$. The authors showed that when the model cluster velocities are convolved with the large observational velocity uncertainties, an artificial high velocity tail, not present in the original model CVF, is produced. Even this artificial high velocity tail was in general not as large as suggested by the data (especially the $D_n - \sigma$ data). Bahcall *et al* suggested that the high velocity tail of the CVF was an artifact of large velocity uncertainties. Differences between the observations and model expectations could arise from underestimated velocity errors. The authors emphasized the need for a cluster sample with higher velocity accuracy in order to better determine the CVF, especially at the critical high velocity end.

Recently, a uniform and accurate sample of peculiar velocities of clusters was obtained by Giovanelli *et al* (1995a,b). Their cluster velocities have considerably greater accuracy than previous studies, mainly due to: (a) access to a homogeneous, all-sky survey, (b) a different TF template relation, based on an extensive study of clusters, (c) an internal extinction correction that allows for larger flux corrections and is luminosity dependent. While the sample size is small (22 groups and clusters out to cz $\leq$ 10 000 km s$^{-1}$), which can thus introduce significant statistical uncertainties, the high quality of the velocity measurements provides a clear advantage. The cluster velocity uncertainties range from $\sim$ 50 km s$^{-1}$ to 340 km s$^{-1}$, with a mean uncertainty of 160 km s$^{-1}$. By contrast, the previous sample uncertainties ranged from $\sim$ 70 km s$^{-1}$ to $\sim$900 km s$^{-1}$, with a mean uncertainty of 410 km s$^{-1}$. We use this sample to determine the CVF and to compare it with model expectations. The reduced observational uncertainties increase the accuracy of the measured CVF, especially at higher velocities.

We present in Figs. 1a–c the CVF determined from the Giovanelli *et al* (1995a,b) sample. (The 1995b sample is slightly larger, with some improvements over the 1995a sample; both overlapping samples yield similar results, and both are presented in Fig. 1a, for comparison.) The error-bars correspond to $\pm\sqrt{N}$ statistical uncertainties. The curves in Fig. 1 represent the CVFs of the four cosmological models (§2.1) *convolved* with the observational velocity uncertainties (for proper comparison with the data). The small velocity uncertainties of this sample have only a minor impact on the true (unconvolved) CVF.

The new data, in contrast with previous samples, do not exhibit a high velocity tail. In fact, there are *no* observed clusters with velocities larger than v$_{1D}$ $\sim$600 km s$^{-1}$, yielding P(v$_{1D}$ > 600 km s$^{-1}$)$\lesssim$ 0.05. In contrast, the previous CVF based on data with larger velocity uncertainties showed a high velocity tail to v$_{1D}$ $\sim$ 2000 km s$^{-1}$, with P(v$_{1D}$> 600 km s$^{-1}$)$\sim$ 0.4, and P(v$_{1D}$> 1000 km s$^{-1}$) $\sim$0.1 (Bahcall *et al* 1994b). Similarly, the root-mean-square velocity of the current cluster sample is $<$ v$^2_{1D}$ $>^{1/2}$ = 293$\pm$28km s$^{-1}$, as compared with 607$\pm$64 km s$^{-1}$ for the previous TF data and 725$\pm$50 km s$^{-1}$ for the previous TF + D$_n\sigma$ data (Bahcall *et al* 1994b).

## 3. Comparison of Models with Observations

The observed and model velocity functions are compared with each other in Figure 1 and Table 2. The main difference among the models is apparent: the $\Omega=1$ models (CDM and HDM, convolved with the observational velocity uncertainties) exhibit a large tail of high velocity clusters, with ~5% of all clusters at $v_{1D} > 1100$ km s$^{-1}$ and 1450 km s$^{-1}$ respectively, while $\Omega = 0.3$ CDM exhibits the lowest cluster velocities, with ~5% of clusters at $v_{1D} > 650$ km s$^{-1}$ (for the convolved models). $\Omega = 0.3$ PBI has intermediate velocities, with ~5% of clusters at $v_{1D} > 900$ km s$^{-1}$. The observed CVF indicates a clear absence of high velocity clusters. This is consistent with the $\Omega \sim 0.3$ CDM model and inconsistent (at $\gtrsim 3\sigma$) with the $\Omega = 1$ CDM and HDM models. We do not observe any clusters with $v_{1D} > 600$ km s$^{-1}$ in this sample; the observed CVF yields P($v_{1D} > 600$ km s$^{-1}$)$\lesssim 0.05$, or $\lesssim 1$ cluster out of a sample of 22 clusters. From the integrated model CVFs (Figs. 1b,c) we would expect *on average* to find 1.7, 4, 6 and 8 clusters with $v_{1D} > 600$ km s$^{-1}$ in a random sample of 22 clusters for $\Omega = 0.3$ CDM, $\Omega = 0.3$ PBI, $\Omega = 1$ CDM, and $\Omega = 1$ HDM respectively (Table 2). The probability that the observed CVF for $v_{1D} > 600$ km s$^{-1}$ is consistent with the various models can be estimated using the binomial distribution statistic, yielding significance levels of 48%, 4%, 1%, and < 0.1%, for $\Omega = 0.3$ CDM, $\Omega = 0.3$ PBI, $\Omega = 1$ CDM, and $\Omega = 1$ HDM respectively. A formal K-S test of the integrated CVF (Fig. 2b) indicates that the data is consistent with the models at significance levels of ~ 90%, 13%, 1% and 0.1% respectively for $\Omega = 0.3$ CDM, $\Omega = 0.3$ PBI, $\Omega = 1$ CDM, and $\Omega = 1$ HDM. A mixed dark matter model, with ~70% CDM and ~30% HDM, is expected to yield results similar to $\Omega \sim 1$ CDM. The results are summarized in Table 2.

The RMS peculiar velocity of clusters in the present sample (§2.2) is compared with model expectations in Table 2. The RMS velocity, as well as the K-S test, and an inspection of Figures 1a–c all suggest that the $\Omega = 0.3$ CDM model is consistent with the observed cluster velocities, while the $\Omega$= 1 models are less consistent with the data. (A PBI model with $\Omega \lesssim 0.3$ is also acceptable). The present sample allows, for the first time, an accurate comparison of cluster velocities with model expectations, not dominated by velocity uncertainties. A larger sample, with similarly accurate velocities, is needed in order to confirm and refine this conclusion.

## 4. Conclusions

We have determined the peculiar velocity function of clusters of galaxies using a small but accurate sample of cluster velocities (Giovanelli *et al* 1995a,b). The relatively accurate velocities enable a reliable determination of the CVF, not dominated by velocity uncertainties. The CVF shows no high velocity clusters, $P(v_{1D} > 600$ km s$^{-1}) \lesssim 0.05$, in contrast with less accurate previous samples that exhibited an artificially large velocity tail to $v_{1D} \sim 2000$ km s$^{-1}$, with $P(v_{1D} > 600$ km s$^{-1}) \sim 0.4$. The root-mean-square 1D cluster velocity is $<v_{1D}^2>^{1/2} = 293 \pm 28$ km s$^{-1}$.

We compare the cluster VF with expectations from several cosmological models. We find the data to be most consistent with a low-density ($\Omega \sim 0.3$) flat CDM model, marginally consistent with a low-density flat PBI model ($\Omega \sim 0.3$), and inconsistent at $\gtrsim 3\sigma$ level with $\Omega$= 1 CDM and HDM models in which a larger high velocity tail is expected. Similarly, the RMS cluster 1D velocities in the models yield (when convolved with the observational uncertainties) 314, 423, 516, and 632 km s$^{-1}$, respectively, as compared with the observed $293 \pm 28$ km s$^{-1}$, further supporting the $\Omega \sim 0.3$ CDM model. A low-density flat CDM model, which best fits other observations, including the mass function and correlation function of clusters (Bahcall & Cen 1992), the baryon density in clusters (White *et al* 1993, Lubin *et al* 1996), the power spectrum and small scale velocities of galaxies (Maddox *et al* 1990, Ostriker 1993), is therefore also consistent with the cluster velocity function.

It is a pleasure to thank Riccardo Giovanelli and Martha Haynes for providing us with their high quality cluster velocity data prior to publication, and for helpful and stimulating discussions. This work is supported by NSF grant AST93-15368.

---

This manuscript was prepared with the AAS LATEX macros v4.0.

## 5. Figure Captions

Fig. 1.— Observed versus model cluster velocity functions. The Giovanelli *et al* (1995a,b) IRTF velocity data(§2.2), is compared with model CVFs convolved with the observational errors. (Due to the small observational errors, the effect of the convolution is small). Fig 1a represents the differential function, Figs. 1b & 1c represent the integrated function on linear and log scales. The solid line histogram (with $\sqrt{N}$ statistical error bars) represents the 22 cluster sample (1995b), while the points (plotted only for Fig. 1a) represent the earlier 16 cluster sample (1995a). Note the absence of a high-velocity tail in the observed CVF.

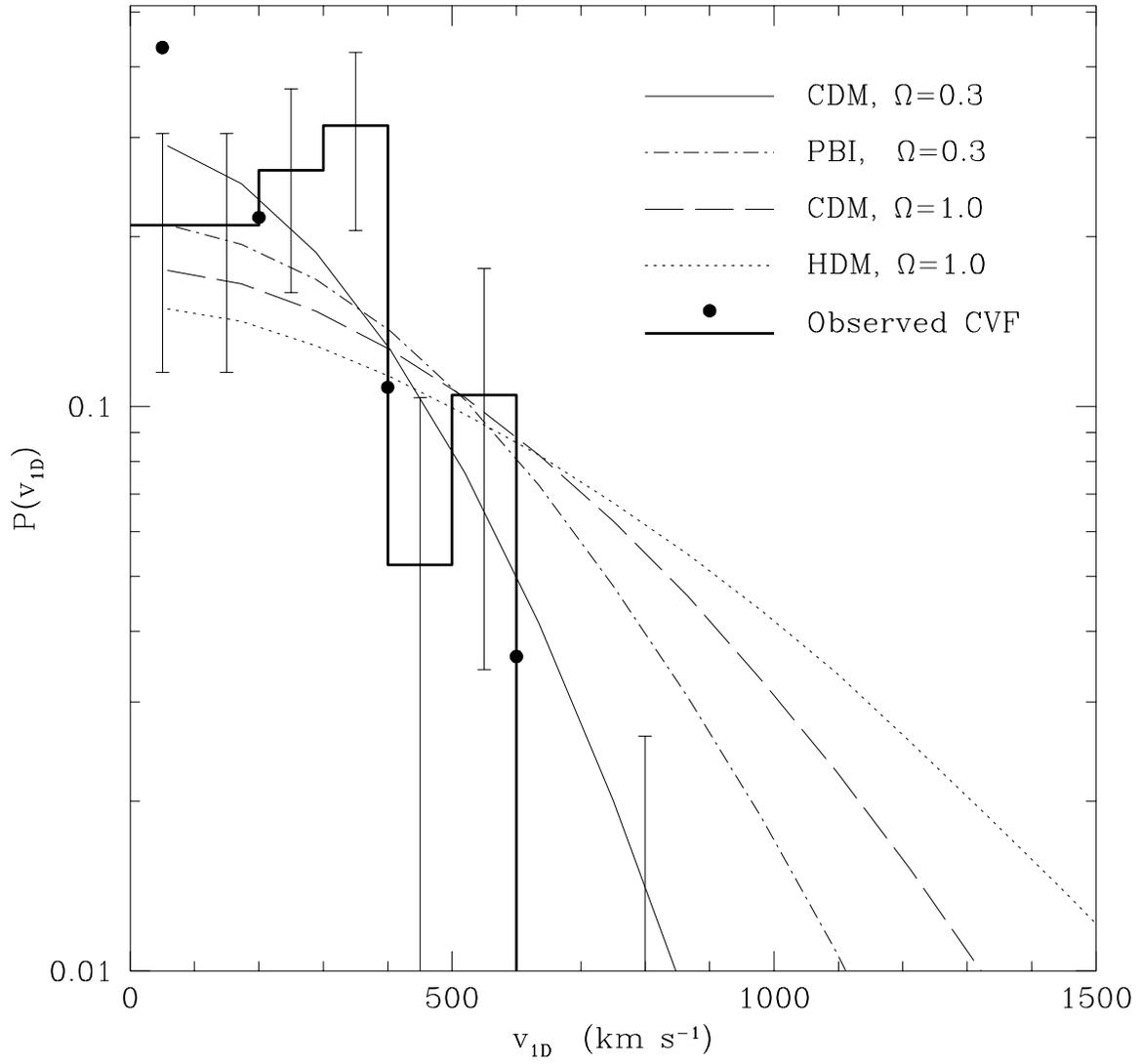

Fig. 1a.—

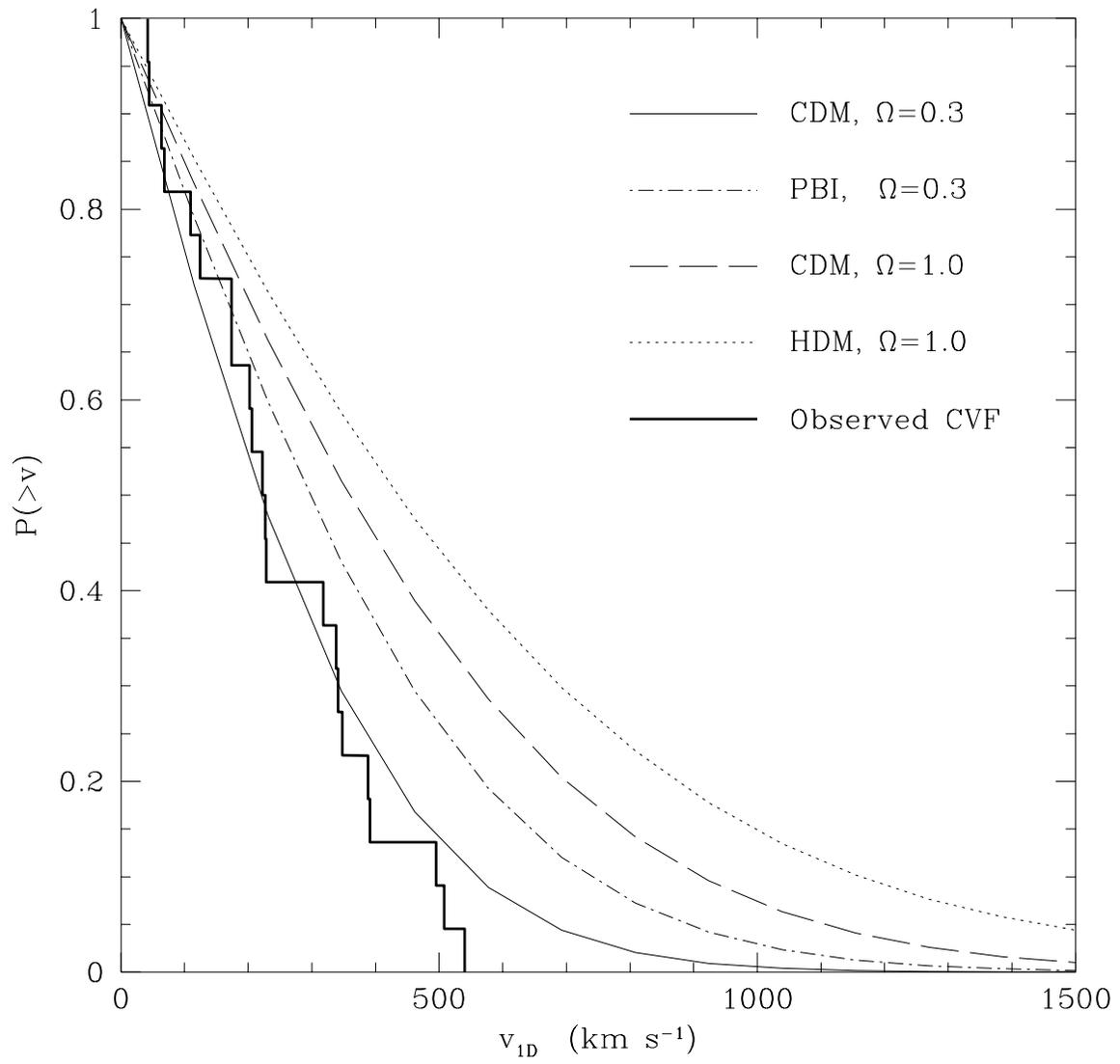

Fig. 1b.—

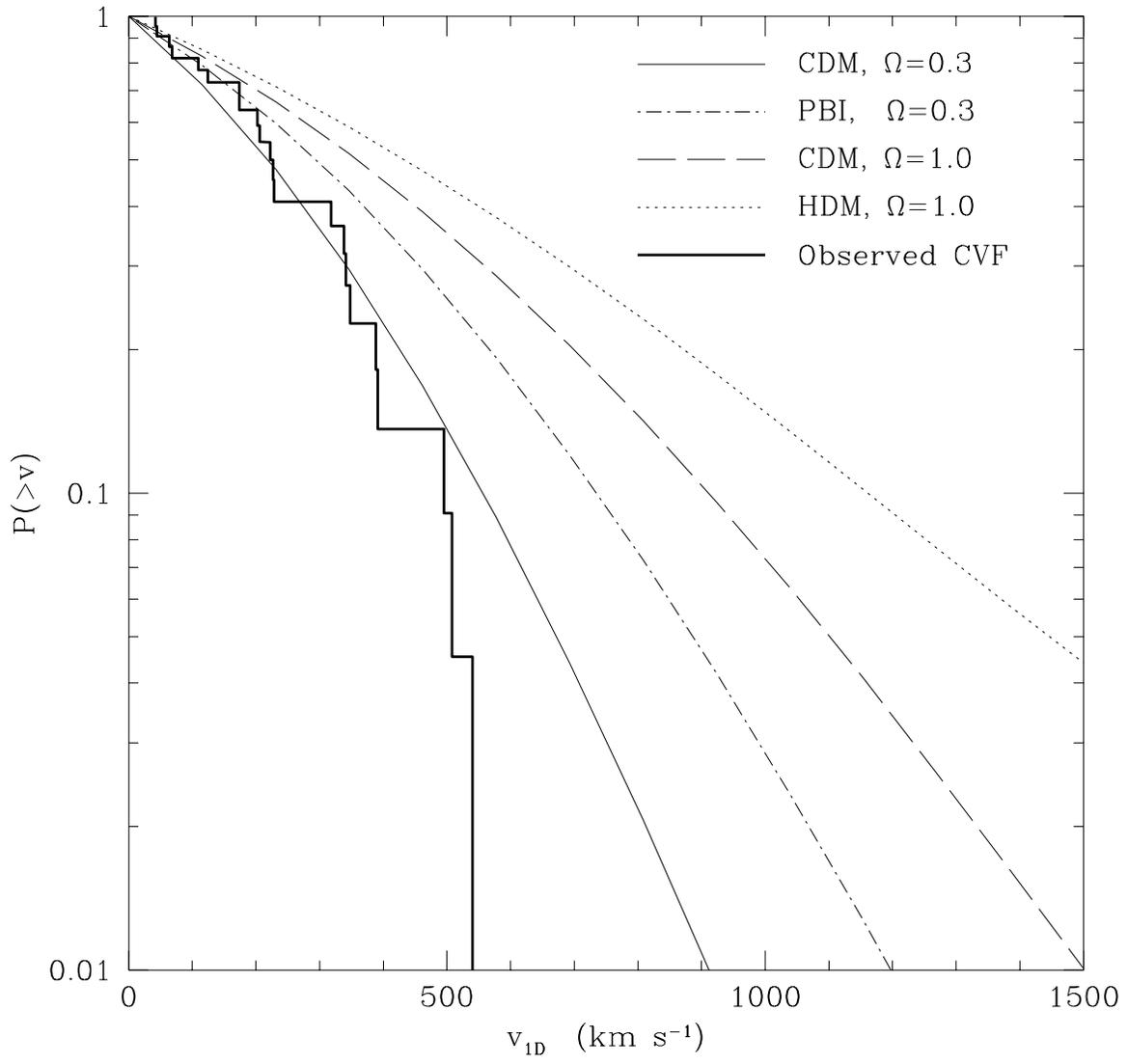

Fig. 1c.—

Table 1.   Model Parameters

| Model | Parameter | | | |
|---|---|---|---|---|
| | $\Omega$ | $\Omega_\Lambda$ | $h$ | $\sigma_8$ |
| CDM | 1.0 | 0.0 | 0.50 | 1.05 |
| CDM | 0.3 | 0.7 | 0.67 | 0.67 |
| HDM | 1.0 | 0.0 | 0.50 | 0.86 |
| PBI | 0.3 | 0.7 | 0.50 | 1.02 |

[a]CDM = Cold Dark Matter model; HDM = Hot Dark Matter model; PBI = Primeval Baryonic Isocurvature model.

Table 2.  Cluster Velocities: Models vs. Observations

|  | (a) $v_{\rm RMS}$ (km s$^{-1}$) | (b) Velocity Function KS test | (c) $N_{\rm cl}$ ($v_{1D} > 600$ km s$^{-1}$) |
|---|---|---|---|
| Observed | $293 \pm 28$ | – | $\leq 1$ |
| CDM $\Omega = 0.3$ | 314 | 90% | 1.7 |
| PBI $\Omega = 0.3$ | 423 | 13% | 4 |
| CDM $\Omega = 1.0$ | 516 | 1% | 6 |
| HDM $\Omega = 1.0$ | 632 | $\leq 0.1\%$ | 8 |

[a]Cluster RMS velocities, $< v_{1D}^2 >^{1/2}$. The model velocities are convolved with the observational velocity uncertainties.

[b]The KS significance levels for the observed vs. model cumulative velocity functions.

[c]The number of observed versus expected clusters with velocities $v_{1D} > 600$km s$^{-1}$, (for a random sample of 22 clusters for each model).